\documentclass{edp-jp4}
\usepackage{graphicx}
\newcommand{\psfigure}[2]{\includegraphics*[width=#1]{#2}}
\begin{document}

%%-----------------------------
%%      the top matter
%%-----------------------------
\title{On the Dynamics and Disentanglement in Thin and \\[1mm]
Two-Dimensional Polymer Films} 
\author{H. Meyer}
\author{T. Kreer}
\author{A. Cavallo}
\author{J. P. Wittmer}
\author{J. Baschnagel}
\address{Institut Charles Sadron, 6 rue Boussingault, 67083 Strasbourg, France}
\maketitle
\begin{abstract} 
We present results from molecular dynamics simulations of strictly two-dimensional 
(2D) polymer melts and thin polymer films in a slit geometry of thickness of the order 
of the radius of gyration.  We find that the dynamics of the 2D melt
is qualitatively different from that of the films.  The 2D monomer mean-square
displacement shows a $t^{8/15}$ power law at intermediate times 
instead of the $t^{1/2}$ law expected from Rouse theory for nonentangled chains.  
In films of finite thickness, chain entanglements may occur.  The impact of confinement 
on the entanglement length $N_\mathrm{e}$ has been analyzed by a primitive path analysis.  
The analysis reveals that $N_\mathrm{e}$ increases strongly with decreasing film 
thickness.
\end{abstract}
%
%%-----------------------------
%%      your text
%%-----------------------------
\section{Introduction}
Thin polymer films are in the focus of current research not only due to their
technological importance, but also because their properties can differ 
substantially from those of the bulk \cite{GranickEtal:JPSB}.  
For example, polymers in the bulk melt are considered ideal Gaussian chains 
\cite{RubinsteinColby}.  Since the extensions of a Gaussian chain in
different spatial directions are uncoupled, Silberberg \cite{Silberberg:Colloid1982}
hypothesized that polymer conformations remain unperturbed in the direction parallel 
to the film surface.  In perpendicular direction, however, conformations may be perturbed 
because chains are folded back into the film at the surface.  This reflection thus leads to 
an increase of the self-density, and one may expect that, for ultrathin films, this increase
could also alter chain conformations and dynamics in direction parallel to the surface.  

The impact of this increased self-density on chain entanglements was discussed
in Ref.~\cite{BrRu96macro} and recent experiments support the idea that the effective 
entanglement density decreases with film thickness \cite{SiMaDVBrJo2005prl}. On the other hand, 
in strictly two dimensions chains do not overlap so that there are no entanglements.  
The properties of two-dimensional (2D) melts have recently been discussed theoretically 
\cite{SeJo2003epje}.  
Interestingly, the theory suggests that the intuitive expectation---if entanglements are absent,
the Rouse model \cite{RubinsteinColby} should apply---is not true.  Chain relaxation 
in 2D is predicted to be {\em faster} than Rouse dynamics.

In this paper, we present results on polymer conformation and dynamics in thin 
films at high temperature (\ie\ far above the glass transition; the dynamics 
close to the glass transition is discussed {\em e.g.}\ in Refs.~\cite{BaVa2005jpcm,PeMeBa2006jpsb}).   
For films of finite thickness, we exemplify the impact of confinement on the entanglement length
for one chain length, $N=256$.  For strictly 2D films, we present data for a broader 
range of chain lengths, which provide evidence for monomer dynamics faster than expected from
Rouse theory, in accord with the predictions of Ref.~\cite{SeJo2003epje}.

\section{Simulation Model}
The results presented in this article are obtained with a flexible, 
purely repulsive bead-spring model.  The model is derived from a coarse-grained
model for polyvinylalcohol (cgpva) \cite{MeMP2001jcp} which has been employed
to study polymer crystallization \cite{crystlong}.  It is characterized by two 
potentials: a harmonic bond potential and a nonbonded potential of Lennard-Jones 
(LJ) type.  The parameters of the bond potential are adjusted so that the ratio of 
bond length $\ell_0$ to monomer diameter $\sigma_0$ is approximately the same as
in the standard Kremer-Grest model (\ie, $\ell_0=0.97\sigma_0$) \cite{KrGr90jcp}.  
While the Kremer-Grest model utilizes a 12-6 LJ potential to describe nonbonded 
interactions, our nonbonded potential has a softer repulsive part.  It is given by
\begin{equation}
U_\mathrm{cgpva}(r) =
 \epsilon_\mathrm{cgpva} \left[\left(\frac{\sigma_0}{{r}}\right)^{9}- 
\left(\frac{\sigma_0}{{r}}\right)^{6} \right] \;,
\label{eq:cgpva}
\end{equation}
where $\epsilon_\mathrm{cgpva}=1.511 k_\mathrm{B}T$ and the monomer diameter $\sigma_0$ is related
to the standard LJ $\sigma$ by $\sigma_0=0.89 \sigma$.  Equation~(\ref{eq:cgpva})
is truncated in the minimum ($r_\mathrm{cut}=1.02\sigma$) and shifted to zero in order
to obtain purely repulsive nonbonded interactions.  In the following we report all data in LJ units: 
lengths are given in units of 
$\sigma$, tempera\-ture in units of $\epsilon/k_\mathrm{B}$ (energy $\epsilon=1$ and
 Boltzmann's constant $k_\mathrm{B}=1$), and time in units of $\tau=\sqrt{\sigma^2 m/\epsilon}$ 
(mass $m=1$).

We perform molecular dynamics simulations in the canonical ensemble with a Langevin 
thermostat (friction constant $\Gamma=0.5$) at temperature $T=1$ and monomer density 
$\rho=1.2$.  (In terms of $\sigma_0$ this density corresponds to $0.84
\sigma_0^{-3}$, the typical melt density of the Kremer-Grest model \cite{KrGr90jcp}.)  
The equations of motion are integrated by the velocity-Verlet algorithm with an 
integration time step of $0.01$.  

Repulsive walls are placed in the $xz$-plane at $y=0$ and $y=D$ by introducing a 
short-range monomer-wall interaction: we choose $U_\mathrm{mw}(y)=2 U_\mathrm{cgpva}(y)$ 
if $y<r_{\rm cut}$ and $D-y<r_{\rm cut}$.  For strictly two-dimensional (2D) films a wall 
distance of $D=1.9$ was chosen.  This distance is so small that the monomers 
only vibrate perpendicular to the film plane with a maximum amplitude of $\pm 0.22$
\cite{Comment_on_crossing}.  For the 2D melt, simulations were also performed with the 
Kremer-Grest model as a ``control experiment''.  The results reported in the next section are 
the same for both models.

\section{Dynamics in strictly two-dimensional films}
In a 2D melt, chains cannot overlap; they must stay separated 
from each other.  These segregated chains have an irregular shape \cite{SeJo2003epje} which
is determined by (binary) contacts with other chains.  Recent theoretical work suggests that
the perimeter $L$ of 2D chains is not proportional to the radius of gyration $R_\mathrm{g}$
(this would be the case if the chains were disk-like objects).  Rather it is predicted that 
$L \sim N^{5/8}$, whereas $R_\mathrm{g} \sim N^{1/2}$ \cite{SeJo2003epje}.  Recent
simulation results support this prediction \cite{CavalloEtal:EPL2003}.  

The theory of Ref.~\cite{SeJo2003epje} considers a chain as sequence of self-similar 
subchains containing $s$ monomers ($1 \ll s \leq N$).  So one expects the radius of gyration 
and the perimeter of the subchains to scale with $s$ as $R_\mathrm{g}(s) \sim s^{1/2}$ and 
$L(s) \sim s^{5/8}$.  These conformational features have an impact on the dynamics.  
Reference~\cite{SeJo2003epje} suggests that the subchain relaxation is controlled by friction 
at chain boundary and predicts that the corresponding relaxation time $\tau(s)$ scales with $s$ 
as $\tau (s) \sim s^{15/8}$ ($1 \ll s \leq N$). 

\begin{figure}
\psfigure{0.5\textwidth}{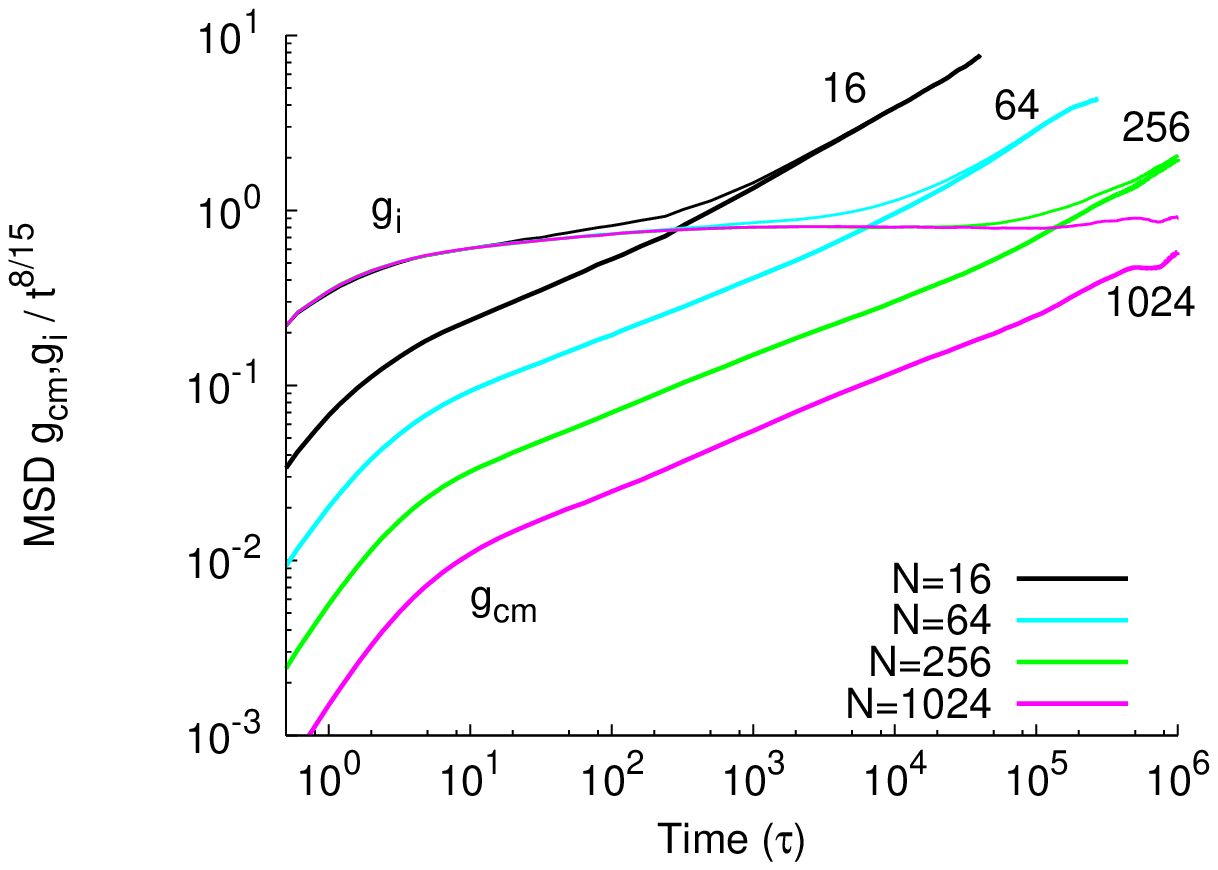}
\hspace{-1mm}
\psfigure{0.5\textwidth}{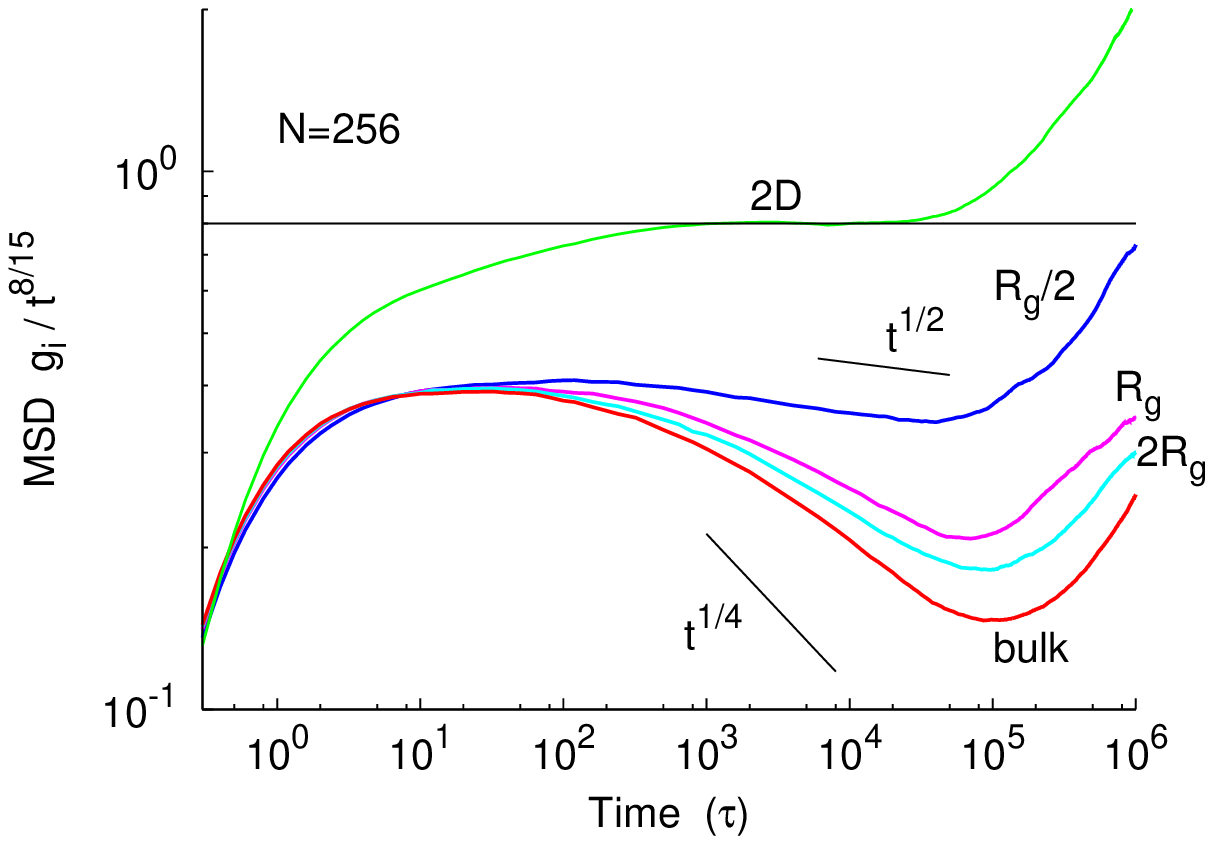}
\caption{Left panel: Mean-square displacements (MSDs) of the center of mass $g_\mathrm{cm}(t)$ and of 
the innermost monomer $g_\mathrm{i}(t)$ in a 2D film.  The MSDs are divided by $t^{8/15}$ (see
text for details). At early times, the monomer displacement is identical for all $N$.  At 
late times, $g_\mathrm{i}(t)$ crosses over to free diffusion, where $g_\mathrm{i}(t) = 
g_\mathrm{cm}(t)$.  The time of this crossover increases with $N$.
Right panel: $g_\mathrm{i}(t)/t^{8/15}$ parallel to the wall versus $t$ for $T=1$ and $N=256$.  
The upper curve corresponds to the 2D case.  The other curves display 
film results for $D=4\approx R_\mathrm{g}^\mathrm{bulk}/2$, $D=8\approx R_\mathrm{g}^\mathrm{bulk}$, 
$D=16\approx 2 R_\mathrm{g}^\mathrm{bulk}$, and the bulk.  The lines labeled by $t^{1/2}$ and $t^{1/4}$
show respectively expectations from Rouse and reptation theories 
\cite{RubinsteinColby}.  The horizontal line indicates the $t^{8/15}$ power law.
For all systems, the densities are equivalent to the 
bulk density ($\rho=1.2$).  According to the primitive path analysis of Sec.~\ref{subsec:ppa}, the bulk 
entanglement length is $N_\mathrm{e} \approx 65$.}
\label{fig:msd2d}
\end{figure}

We can test this prediction by analyzing the mean-square displacement (MSD) of the 
innermost monomer of a chain, $g_\mathrm{i}(t)$.  For displacements inside the volume 
pervaded by a chain, the distance covered by a monomer in time $t$ should be proportional to 
$R_\mathrm{g}(s)$.  Thus, we expect to find $g_\mathrm{i}(t) \sim R_\mathrm{g}^2(s) \sim s \sim 
t^{8/15}$, instead of the Rouse prediction $g_\mathrm{i}(t) \sim t^{1/2}$ \cite{RubinsteinColby}.  
Figure~\ref{fig:msd2d} shows that this prediction agrees well with the simulation.  Moreover,
the right panel of Fig.~\ref{fig:msd2d} reveals that the simulated 
$g_\mathrm{i}(t)$ is sufficiently precise to distinguish between $t^{8/15}$ and 
$t^{1/2}$---albeit the exponents are numerically very close---and that Rouse behavior may be 
ruled out.    

In addition to that, we find that also the center of mass exhibits a power-law regime where 
$g_\mathrm{cm}(t) \sim t^{\approx 0.87}$ before free diffusion sets in.  This power law, however, is 
not yet understood.

\section{Chain conformations and entanglements in thin films}
In this section we present simulation results for polymer films having a thickness $h$
($=D-\sigma_0$) that exceeds the monomer diameter sufficiently so that chains can overlap.  All films 
contain 48 chains of length $N=256$.  They were prepared by slow compression of bulk configurations 
followed by long equilibration at constant $h$ (equilibration time $>3\times 10^5$; the end-to-end 
vector autocorrelation function decays to 0.1 in $\tau_\mathrm{ee}\approx 1.5\times10^5$).  In the 
following, we first discuss the dependence of the chain size on film thickness.  This analysis will 
suggest that finite-$h$ effects become prominent for $D=4,8,16$ ($R_\mathrm{g}^\mathrm{bulk}=
7.5$).  Accordingly, we focus on these film thicknesses in our subsequent discussion of polymer 
dynamics.

\subsection{Thickness dependence of the chain size}
\begin{figure}
\psfigure{0.5\textwidth}{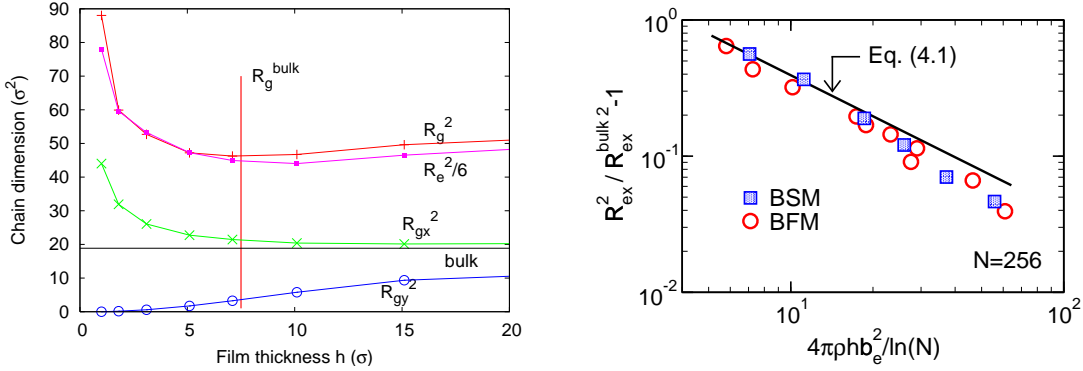}
\hfill
\psfigure{0.45\textwidth}{fig2b.eps}
\caption{Left panel: Chain dimensions versus $h$ for $N=256$.  $R_{\mathrm{g}x}$ and
$R_{\mathrm{g}y}$ are respectively the parallel ($x$) and perpendicular ($y$) components of 
$R_\mathrm{g}$ in the films. $R_\mathrm{e}$ is the end-to-end distance of a chain.  
The bulk result ($R_{\mathrm{g}x}^2= R^{\mathrm{bulk}\,2}_\mathrm{g}/3\simeq 18.75$) is  
shown by a horizontal line.  The vertical line indicates $h=R^{\mathrm{bulk}}_\mathrm{g} = 7.5$. 
Right panel: Scaling plot according to Eq.~(\ref{eq:reethinfilms}) 
for the present bead-spring model (BSM: $b_\mathrm{e}=1.17$, $\rho=1.2$) and the bond-fluctuation 
lattice model (BFM: $N=256$, $b_\mathrm{e}=3.2$, $\rho=1/16$).  The BFM data are taken from 
Ref.~\cite{CaMuWiJoBi2005jpcm}.}
\label{fig:rg-filmD}
\end{figure}

Figure~\ref{fig:rg-filmD} shows the components of the radius of gyration parallel ($R_{\mathrm{g}x}$) 
and perpendicular ($R_{\mathrm{g}y}$) to the wall as a function of film thickness $h$.  For $h > 
R_\mathrm{g}^\mathrm{bulk}$ we find that the parallel component remains very close to the bulk value, 
\ie\ $R_{\mathrm{g}x}^2= R^{\mathrm{bulk}\,2}_\mathrm{g}/3$, whereas the perpendicular
component decreases strongly.  This behavior agrees well with results from other
simulations \cite{Mul2002jcp,CaMuWiJoBi2005jpcm} and experiments (measuring $R_{\mathrm{g}x}$) 
\cite{JonesEtal:Nature1999}.  It may be interpreted, to first approximation, as an evidence for 
ideal chain behavior in 
sufficiently thick polymer films.  The term ``ideal behavior'' means that chain conformations are 
random-walk like \cite{RubinsteinColby}.  Since the dimensions of a random walk in $x$, 
$y$ and $z$ directions are uncoupled, Silberberg suggested that in polymer films only the chain 
extension perpendicular to the film surface is depressed below its bulk value by the finite film 
thickness, whereas the parallel component remains unaltered (``Silberberg hypothesis'') 
\cite{Silberberg:Colloid1982}.   

Figure~\ref{fig:rg-filmD} shows that Silberberg's hypothesis, albeit approximately valid for 
$h > R_\mathrm{g}^\mathrm{bulk}$, becomes violated on decreasing $h$ below $R_\mathrm{g}^\mathrm{bulk}$:
we find that $R_{\mathrm{g}x}> R_{\mathrm{g}x}^\mathrm{bulk}$.  This systematic trend of chain 
swelling suggests that the screening of intrachain excluded volume interactions, responsible for 
ideal chain behavior \cite{WittmerPRL2004}, progressively diminishes with decreasing $h$.  

Qualitatively, this deviation from ideal behavior may be rationalized as follows
\cite{Mul2002jcp,CaMuWiJoBi2005jpcm}.  Screening of intrachain excluded volume interactions in the 
bulk is caused by strong chain overlap.  Since the monomer density 
$\rho_\mathrm{c}$ resulting from the $N$ monomers of the chain in its pervaded volume 
($\sim R_\mathrm{g}^{\mathrm{bulk}\,3}$) is very small ($\rho_\mathrm{c} = N/
R_\mathrm{g}^{\mathrm{bulk}\,3} \sim N^{-1/2}$), a chain must have $N^{1/2}$ contacts with 
other chains to maintain the constant melt density $\rho$.  Thus, a condition for strong chain 
overlap is $\rho_\mathrm{c}/\rho =  N/(\rho R_\mathrm{g}^{\mathrm{bulk}\,3}) \ll 1$.  When 
confining the melt to thicknesses $h \leq R_\mathrm{g}^\mathrm{bulk}$, $\rho_\mathrm{c}$
changes: $\rho_\mathrm{c}(h) = N/hR_\mathrm{g}^{\mathrm{bulk}\,2}$.  Thus,  with decreasing film 
thickness it becomes progressively more difficult to fulfil the condition $\rho_\mathrm{c}/\rho \ll 1$.

The impact of this reduced overlap on chain conformations in ultrathin polymer films has 
recently been worked out by Semenov and Johner \cite{SeJo2003epje}.  Their theory predicts 
strong devia\-tions from ideal behavior.  For the parallel component of the end-to-end distance, 
$R_\mathrm{e\|}^2(h,N)$ ($=R_{\mathrm{e}x}^2+R_{\mathrm{e}z}^2$), the result reads 
\begin{equation}
R^2_\mathrm{e\|}(h,N) \simeq N b^2_\mathrm{e} \bigg [ 1 + f(h) \frac{4}{b^2_\mathrm{e}} \ln N \bigg ], 
\quad f(h) = \frac{1}{4\pi\rho h} \;,
\label{eq:reethinfilms}
\end{equation}
where $b_\mathrm{e}$ is (taken to be) the statistical segment length of the bulk.

Figure~\ref{fig:rg-filmD} tests this prediction.  It depicts a scaling of $R_{\mathrm{e}x}(h)$
according to Eq.~(\ref{eq:reethinfilms}) for two data sets, the results obtained for the present 
bead-spring model and the results published in Ref.~\cite{CaMuWiJoBi2005jpcm} for the bond-fluctuation 
lattice model.  Within the statistical uncertainties the simulation results for both (microscopically 
very distinct) models agree with one another and with the theoretical prediction of 
Eq.~(\ref{eq:reethinfilms}).

\subsection{Impact of film thickness on the entanglement density}
\label{subsec:ppa}
The right panel of Fig.~\ref{fig:msd2d} compares $g_\mathrm{i}(t)$ in 2D with that for films
of finite thickness (in the latter case, $g_\mathrm{i}(t)$ is measured parallel to 
the walls).  While monomer motion is faster than Rouse dynamics in 2D, we find a slowing down of the 
MSD at intermediate times for finite film thickness, which increases with increasing $h$.  
We speculate that
this systematic trend towards {\em slower} monomer motion is attributed to an {\em increase} of the 
entanglement density when the film thickness approaches bulk-like values.

\begin{figure}
\centerline{
\psfigure{0.26\textwidth}{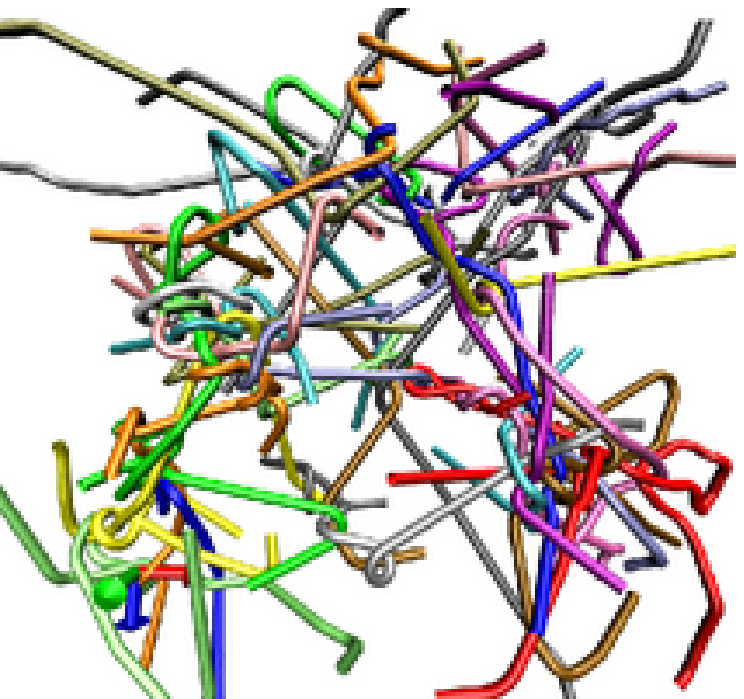}\quad
\psfigure{0.3\textwidth}{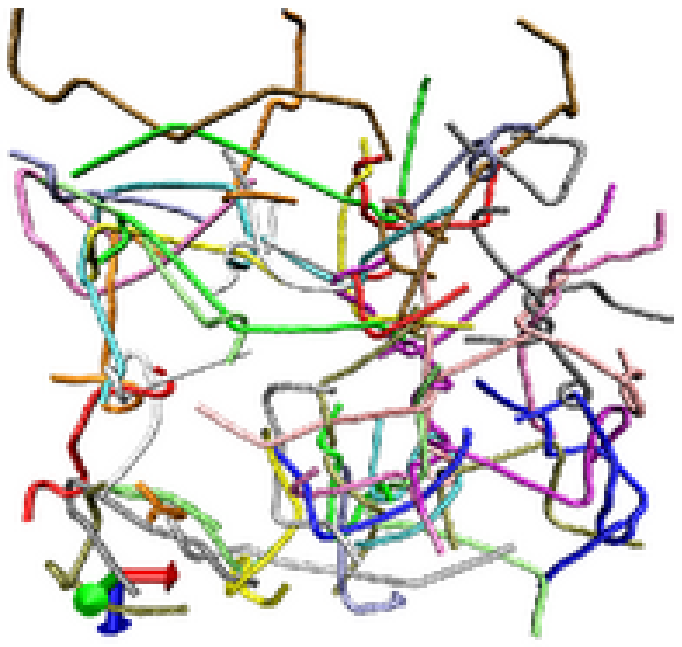}\quad
\psfigure{0.4\textwidth}{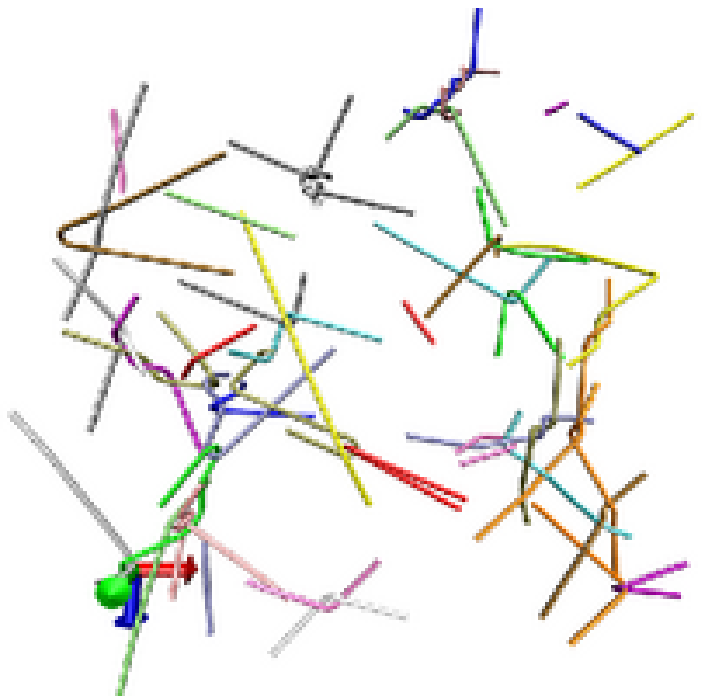}}
\caption{Primitive paths determined by the algorithm of Ref.~\cite{EvSuGrEA2004sci}.  
View along the normal of the film; laterally, periodic boundary conditions apply. 
Left: $D=16\approx 2R_\mathrm{g}$, center: $D=8\approx R_\mathrm{g}$, 
right: $D=4\approx 0.5R_\mathrm{g}$.  For the thinner films, the entanglement network
is not percolating any more.}
\label{fig:ppa}
\end{figure}

To test this idea we performed a primitive path analysis (PPA) via the algorithm proposed in
Ref.~\cite{EvSuGrEA2004sci,SiMaDVBrJo2005prl}.  This algorithm consists in contracting
the bonds of each chain while fixing the chain ends and preserving the
excluded volume only between monomers of different chains. 
The resulting network of primitive paths is shown in Fig.~\ref{fig:ppa} for three choices
of $D$.  From the statistics of
the primitive path network one can directly deduce the entanglement
length $N_\mathrm{e}$ due to interchain entanglements \cite{EvSuGrEA2004sci}. 
We performed this analysis for at least 5 configurations for each film thickness. 
Considerable fluctuations were found which are probably due to the small number of chains
in the simulation box. 
However, we checked that these fluctuations are not due to finite size effects: Doubling of the 
box size for a given configuration leaves the results obtained from the PPA 
almost unchanged. 

A clear trend emerges from Fig.~\ref{fig:ppa}.  While for $D=2R_\mathrm{g}^\mathrm{bulk}$ 
the entanglement length, $N_\mathrm{e}=72\pm 6$, is only slightly larger than the bulk 
value ($N_\mathrm{e}=68\pm 6$), smaller film thicknesses lead to strong 
disentanglement: $N_\mathrm{e}=90\pm 8$ for $D=R_\mathrm{g}^\mathrm{bulk}$ and 
$N_\mathrm{e}=135\pm 10$ for $D=R_\mathrm{g}^\mathrm{bulk}/2$ \cite{Comment_on_entanglements}.  
The thinnest film may be considered as practically disentangled.
This conclusion is supported, on the one hand, by the snapshot of Fig.~\ref{fig:ppa}---the 
primitive path mesh does not percolate---and, on the other hand, by the MSD which shows 
no slowing down with respect to Rouse dynamics.

\section{Conclusion}
The presented simulation results show that 2D polymer melts display monomer dynamics that is faster 
than expected from Rouse theory.  We find no indication of a reptation-like slowing down, even not
for $N=1024$ which shows the signature of reptation dynamics in the bulk.  With increasing film 
thickness, however, the effect of entanglements on the dynamics becomes gradually
visible.  This implies that $N_\mathrm{e}$ is altered by spatial confinement; 
in ultrathin films it increases with decreasing $h$.  The primitive path analysis of 
Ref.~\cite{EvSuGrEA2004sci}---determining interchain entanglements---allows 
to quantify the extent of disentanglement with respect to the bulk.

An issue for future studies is the role of intrachain entanglements. The model underlying the
analysis of Ref.~\cite{SiMaDVBrJo2005prl} assumes that the total entanglement density remains 
constant. While the number of interchain entanglements is supposed to
decrease with film thickness, the number of intrachain entanglements should 
increase.  The PPA algorithm can be modified such that intrachain entanglements are
conserved beyond a certain chemical distance \cite{SuGrKrEv2005jpsb}.  Preliminary tests suggest
that when preserving intrachain entanglements beyond 64 monomers, almost no difference is
found.  When decreasing this distance to 16 neighboring monomers, the entanglement
lengths are smaller than the values reported here, and the effect appears to be stronger in
thin film.  However, we presently feel that the inclusion of intrachain entanglements cannot 
compensate the loss of interchain entanglements discussed above.  We thus suggest that the 
{\em total} entanglement density decreases.  A more detailed analysis of this problem is
underway.

\begin{acknowledgements}
We thank A. Johner and K. Binder for numerous helpful discussions.  We are indebted to the DFG 
(KR 2854/1-2), the Universit\'e Louis Pasteur, the IUF, and the ESF STIPOMAT programme 
for financial support.  A generous grant of computer time by the IDRIS (Orsay) is also gratefully 
acknowledged.
\end{acknowledgements}

%%-----------------------------
%%      your bibliography
%%-----------------------------
% \bibliographystyle{prsty}
% \bibliography{/home/users/hmeyer/WORK/visc/confit2006/bib/divpol,/home/users/hmeyer/WORK/visc/confit2006/bib/cg,/home/users/hmeyer/WORK/visc/confit2006/bib/crystpol,/home/users/baschnag/TEX/paper/BibTex_Files/references_Polymer.bib,/home/users/baschnag/TEX/paper/BibTex_Files/references_Films.bib}

\begin{thebibliography}{10}
\bibitem{Comment_on_crossing}
The energy cost for two polymers to cross each other is $E_\mathrm{x} \approx 70 k_\mathrm{B}
T$ if $D=1.9$.  By contrast, we found a maximum energy of $E_\mathrm{c} \approx 8 
k_\mathrm{B} T$ for the closest approach between two monomers in the simulation.  
Since $E_\mathrm{c} \ll E_\mathrm{x}$, no chain crossings were observed; the film with $D=1.9$ 
may be considered as ``strictly'' two-dimensional.
\bibitem{Comment_on_entanglements}
We feel that our results should not be interpreted in the sense that the criterion for
an increase of $N_\mathrm{e}$ is necessarily $h < {R_\mathrm{g}}^\mathrm{bulk}$.  
It may be that it rather requires $h$ to become smaller than the tube 
diameter ${d_\mathrm{t}}^\mathrm{bulk}$.  A test of this hypothesis necessitates  
simulations of larger $N$ because the two scales, ${d_\mathrm{t}}^\mathrm{bulk}$ and
${R_\mathrm{g}}^\mathrm{bulk}$, are not well separated for the studied chain length 
$N=256$.
\end{thebibliography}

\end{document}